\documentclass[preprint,showpacs,preprintnumbers,amsmath,amssymb]{revtex4}

\usepackage{graphicx}
\usepackage{dcolumn}
\usepackage{bm}

\begin{document}    
\title{Evolution of the electronic structure across the filling-control and bandwidth-control metal-insulator transitions in pyrochlore-type Ru oxides}
\author{J. Okamoto\cite{address1}, S.-I. Fujimori and T. Okane}
\affiliation{Synchrotron Radiation Research Center, Japan Atomic Energy Research Institute, SPring-8, Sayo-gun, Hyogo 679-5148, Japan}
\author{A. Fujimori}
\affiliation{Department of Complexity Science 
and Engineering, University of Tokyo, Kashiwa, Chiba 277-8561, Japan and Synchrotron Radiation Research Center, Japan Atomic Energy Research Institute, SPring-8, Sayo-gun, Hyogo 679-5148, Japan}
\author{M. Abbate}
\affiliation{Departamento de F$\acute{\rm i}$sica, Universidade Federal de Paran$\acute{\rm a}$, Caixa Postal 19091, Curitiba PR 81531-990, Brazil}
\author{S. Yoshii\cite{Yaddress1} and M. Sato}
\affiliation{Department of Physics, Nagoya University, Chikusa-ku, Nagoya 464-8602, Japan}
\date{\today}

\begin{abstract}  
We have performed photoemission and soft x-ray absorption studies of pyrochlore-type Ru oxides, namely, the filling-control system Sm$_{2-x}$Ca$_x$Ru$_2$O$_7$ and the bandwidth-control system Sm$_{2-x}$Bi$_x$Ru$_2$O$_7$, 
which show insulator-to-metal transition with increasing Ca and Bi concentration, respectively. Core levels and the O 2$p$ valence band in Sm$_{2-x}$Ca$_x$Ru$_2$O$_7$ show almost the same amount of monotonous upward energy shifts with Ca concentration, which indicates that the chemical potential is shifted downward due to hole doping. The Ru 4$d$ band in Sm$_{2-x}$Ca$_x$Ru$_2$O$_7$ is also shifted toward the Fermi level ($E_F$) with hole doping and the density of states (DOS) at $E_F$ increases. The core levels in Sm$_{2-x}$Bi$_x$Ru$_2$O$_7$, on the other hand, do not show clear energy shifts except for the Ru 3$d$ core level, whose line shape change also reflects the increase of metallic screening with Bi concentration. We observe pronounced spectral weight transfer from the incoherent to the coherent parts of the Ru $4d$ $t_{2g}$ band with Bi concentration, which is expected for a bandwidth-control Mott-Hubbard system. The increase of the DOS at $E_F$ is more abrupt in the bandwidth-control Sm$_{2-x}$Bi$_x$Ru$_2$O$_7$ than in the filling-control Sm$_{2-x}$Ca$_x$Ru$_2$O$_7$, in accordance with a recent theoretical prediction. Effects of charge transfer between the Bi 6$sp$ band and the Ru 4$d$ band are also discussed. 
\end{abstract}
\pacs{71.30.+h, 71.27.+a, 71.28.+d, 79.60.-i}
\maketitle

\section{Introduction}
There are two types of metal-insulator transitions (MIT's) in Mott-Hubbard systems, that is, filling-control and bandwidth-control MIT's.\cite{cit4} A fundamental problem is what are common features and what are different features between the two types of transitions. From the spectroscopic point of view, Zhang $et$ $al$.\cite{Zhang1} and Kajueter $et$ $al$.\cite{Kajueter1} have made a remarkable prediction of spectral weight transfer based on dynamical mean-field theory as shown in Fig. \ref{Mott}. In the bandwidth-control system, as $U$/$D$ decreases from the bottom (insulating) to top (metallic) of Fig. \ref{Mott} (b), where $U$ is the on-site Coulomb energy and $D$ is the half bandwidth, spectral weight is transferred from the incoherent part (upper and lower Hubbard bands) toward the coherent part (quasiparticle peak) around the Fermi level ($E_F$). In the filling-control system, the intensity of the coherent part increases with hole concentration $\delta$ with an overall energy shift, reflecting the chemical potential shift, as shown in Fig. \ref{Mott} (a) [for the case of $U$/$D$ = 4].

As for actual Mott-Hubbard systems, it has been reported that the photoemission spectra of Ca$_{1-x}$Sr$_x$VO$_3$\cite{SCVO3S,SCVO3B} and Ca$_{1-x}$Sr$_x$RuO$_3$\cite{CSRO3T} show changes which reflect the bandwidth-control MIT. Unfortunately, Ca$_{1-x}$Sr$_x$VO$_3$ and Ca$_{1-x}$Sr$_x$RuO$_3$ do not undergo an MIT and therefore the vicinity of MIT cannot be investigated. La$_{1-x}$Sr$_x$TiO$_3$ \cite{LSTO3,Fujimori1} and La$_{2-x}$Sr$_x$CuO$_4$ \cite{LSCO4} are well known filling-control systems and have been studied by photoemission spectroscopy.

In order to address the issue of the differences between filling-control and bandwidth-control MIT's, one would like to have a Mott-Hubbard system in which both types of transitions are realized. However, there are few such systems, and pyrochlore-type Ru oxides $A_{2}$Ru$_2$O$_7$ is one of such rare systems. Sm$_{2-x}$Ca$_x$Ru$_2$O$_7$ ($0 \leq x \leq 0.6$) shows an insulator-to-metal transition at $x$ $\sim$ 0.45 with increasing Ca concentration, i.e., through hole doping into the Ru 4$d$ $t_{2g}$ band of the Mott insulator Sm$_2$Ru$_2$O$_7$.\cite{cit3} Sm$_{2-x}$Bi$_x$Ru$_2$O$_7$ ($0 \leq x \leq 2.0$) exhibits an insulator-to-metal transition at $x$ $\sim$ 0.6 with increasing Bi concentration, i.e., through increasing the Ru 4$d$ $t_{2g}$ band width.\cite{cit1,cit2} In Sm$_{2-x}$Bi$_x$Ru$_2$O$_7$, the Ru-O-Ru bond angle changes from 132$^{\circ}$(for $x$ = 0) to 139$^{\circ}$($x$ = 2). Pyrochlore-type transition-metal oxides are composed of $M$O$_6$ octahedra as in the case of perovskite-type transition-metal oxides. The transition-metal $d$-band width, which is significantly affected by the overlap between the transition-metal $d$ orbitals via the ligand O 2$p$ orbitals, is narrower in the pyrochlores than that in the perovskites because the $M$-O-$M$ bond angle $\sim$ 135$^{\circ}$ in the pyrochlores is smaller than 155$^{\circ}$-180$^{\circ}$ in the perovskites.\cite{cit0} Therefore, electron correlation within the $d$ band is expected to be stronger in the pyrochlores than in the perovskites. As for the spectral intensity in the vicinity of the Fermi level, it has been reported that the valence-band ultra-violet photoemission spectra of Sm$_{2-x}$Ca$_x$Ru$_2$O$_7$ and Sm$_{2-x}$Bi$_x$Ru$_2$O$_7$ reflect the doping dependence of their transport properties.\cite{okamoto1} But the difference of the electronic structures and spectral shapes owing to the difference of the electron correlation and the MIT mechanism are still unclear.

In the present work, we have studied core-level energy shifts by x-ray photoemission spectroscopy (XPS) and spectral changes in the valence band by ultra-violet photoemission spectroscopy (UPS) and x-ray absorption spectroscopy (XAS) of Sm$_{2-x}$Bi$_x$Ru$_2$O$_7$ and Sm$_{2-x}$Ca$_x$Ru$_2$O$_7$ across the metal-insulator transitions. From comparison of the XPS and UPS spectra of Sm$_{2-x}$Bi$_x$Ru$_2$O$_7$ and those of Sm$_{2-x}$Ca$_x$Ru$_2$O$_7$, we discuss similarities and differences between the two types of MIT's.

\section{Experimental}
Sintered polycrystalline samples of Sm$_{2-x}$Ca$_x$Ru$_2$O$_7$ and 
Sm$_{2-x}$Bi$_x$Ru$_2$O$_7$ were synthesized by solid-state 
reaction.\cite{cit2,cit3} 
XPS measurements were performed in an ultra high vacuum of $\sim 10^{-10}$ Torr by using a Mg x-ray source ($h\nu$ = 1253.6 eV) with the total energy resolution of 0.8 eV. Energy calibration was achieved by measuring the Au 4$f_{7/2}$ peak (binding energy $E_B$ = 83.9 eV) and the Fermi edge of Au evaporated on the samples. We cleaned the sample surfaces by scraping with a diamond file in the ultra high vacuum of the spectrometer before each measurement. We checked the surface condition by monitoring the high-binding energy shoulder of the O 1$s$ peak, which originates from contaminations or degradation and could not be completely removed by scraping. UPS measurements were done in an ultra high vacuum of $\sim 5 \times 10^{-10}$ Torr using a He discharge lamp ($h\nu$ = 21.2 eV). Energy resolution was 30 meV and 5 meV for UPS measurements of Sm$_{2-x}$Ca$_x$Ru$_2$O$_7$ and Sm$_{2-x}$Bi$_x$Ru$_2$O$_7$, respectively. UPS spectra were taken at liquid nitrogen temperature for Sm$_2$Ru$_2$O$_7$ to avoid charging effect, and at 14$-$26 K for the other samples. We checked the surface condition by monitoring the emission at $\sim$ 10 eV below the Fermi level, which is known to originate from surface contamination or degradation. XAS measurements were done at the spherical-grating-monochromator beamline of the National Synchrotron Light Laboratory in Campinas. Energy resolution was set to better than 0.5 eV and the pressure was below 1$\times$ 10$^{-9}$ Torr. O 1$s$ XAS spectra were taken at room temperature.

\section{Results and discussions}
\subsection{Sm$_{2-x}$Ca$_x$Ru$_2$O$_7$}
Figure \ref{CaXPS} shows the valence-band UPS spectra, the Ru 3$d$, O 1$s$ and Sm 3$d_{5/2}$ core-level XPS spectra, and the O 1$s$ XAS spectra of Sm$_{2-x}$Ca$_x$Ru$_2$O$_7$ ($0 \leq x \leq 0.6$). In the valence-band UPS spectra [Fig. \ref{CaXPS} (a)], the emission from the Fermi level ($E_F$) to $\sim$ 2.5 eV below it mainly consists of Ru 4$d$ $t_{2g}$ states and the emission from $\sim$ 2.5 eV to $\sim$ 10 eV below $E_F$ of O 2$p$ states. Vertical lines for the Ru 3$d$ [Fig. \ref{CaXPS} (b)], O 1$s$ [Fig. \ref{CaXPS} (c)] and Sm 3$d_{5/2}$ [Fig. \ref{CaXPS} (d)] core-level XPS spectra show the energy position of the half maximum of each peak. In the valence-band UPS spectra [Fig. \ref{CaXPS} (a)], the vertical line is set at the half maximum on the lower binding energy side of the O 2$p$ band. With increasing Ca concentration, the vertical lines in the Ru 3$d$, O 1$s$, Sm 3$d_{5/2}$ and O 2$p$ spectra are monotonously shifted toward lower binding energies. As shown in Fig. \ref{CaXPS} (e), the energy shifts of the four spectra are the same within the experimental uncertainties, which means that a rigid band-like energy shift occurs in Sm$_{2-x}$Ca$_x$Ru$_2$O$_7$ due to the chemical potential shift. The O 1$s$ XAS spectra [Fig. \ref{CaXPS} (f)] show unoccupied states hybridized with O 2$p$ states: Ru 4$d$ $t_{2g}$ and $e_g$ bands are seen in the region of 528$-$530 eV and 530$-$534 eV, respectively, as in the inverse photoemission spectrum of Tl$_2$Ru$_2$O$_7$.\cite{Tl2Ru2O7} The leading peak at $\sim$ 529 eV shows a doping-induced shift toward lower energies, which is also consistent with the chemical potential shift. Also, the intensity of the peak increases with Ca substitution as shown in Fig. \ref{CaXPS} (g), consistent with hole doping.

Figure \ref{CaUPS} (a) shows an enlarged plot of the valence-band UPS spectra of Sm$_{2-x}$Ca$_x$Ru$_2$O$_7$ near $E_F$ across the metal-insulator transition, i.e., in the Ru $4d$ $t_{2g}$ band region.\cite{okamoto1} With increasing Ca concentration, the broad peak at $\sim$ 1.5 eV is shifted toward $E_F$ (like the core and valence levels) and thus the spectral intensity around $E_F$ increases. 
A clear Fermi edge is established at $x$ $\sim$ 0.4. As shown in Fig. \ref{CaUPS} (b), the spectral intensity at or around $E_F$ rapidly increase around x $\sim$ 0.4, but the increase becomes weaker for $x \geq$ 0.45. This corresponds well to the transport properties of Sm$_{2-x}$Ca$_x$Ru$_2$O$_7$ that the most dramatic decrease of the electrical resistivity occurs at $x_t \sim$ 0.45.\cite{cit3} The O $2p$ band edge is also shifted toward $E_F$. Thus, the shift of the valence band can be well explained by the chemical potential shift caused by the hole doping into the Ru $4d$ $t_{2g}$ band, consistent with the core-level results. 

Figure \ref{RefShift} shows the UPS spectra of Sm$_{2-x}$Ca$_x$Ru$_2$O$_7$ arbitrarily shifted so that the spectra at and just below $E_F$ of the shifted spectra coincide with that of the metallic end ($x$ = 0.6). The shifted spectra almost coincide with each other except for the slight decrease of the intensity at 0.5 eV with $x$. This reduced spectral weight should be transferred above $E_F$ corresponding to the decrease in the filling of the Ru 4$d$ band.

\subsection{Sm$_{2-x}$Bi$_x$Ru$_2$O$_7$}
In the case of the bandwidth-control Sm$_{2-x}$Bi$_x$Ru$_2$O$_7$ (0 $\leq x \leq$ 2), on the other hand, spectral changes across the metal-insulator transition are more complicated. Figure \ref{EshiftBi} shows the valence-band UPS spectra, the Ru 3$d$, O 1$s$, Bi 4$f$ and Sm 3$d_{5/2}$ core-level XPS spectra and the O 1$s$ XAS spectrum of Sm$_{2-x}$Bi$_x$Ru$_2$O$_7$ ($0 \leq x \leq 2$). 
Vertical line in each spectrum shows the energy position of the half-maximum of each peak. Figure \ref{EshiftBi} (f) summarizes the energy shift of the vertical lines. As the 
Bi concentration increases, the Ru 3$d$ level (and O 2$p$ level) shows a monotonous shift toward lower binding energies, while the other core levels show small and nonmonotonous shifts. They are shifted toward lower binding energies for Bi concentration $x \leq 0.6$ but slightly toward higher binding energies for Bi concentration $x \geq$ 0.6. The shift of the O 2$p$ level toward lower binding energies is also weakened at Bi concentration $x \geq$ 1.2. The upward shift of the core levels and the O 2$p$ band for small $x$ may indicate that there is partial hole doping into the Ru 4$d$ $t_{2g}$ band from the Bi 6$sp$ band, as suggested by the recent optical studies of Y$_{2-x}$Bi$_x$Ru$_2$O$_7$.\cite{Noh} The line shape of the Bi 4f$_{7/2}$ changes from symmetric to asymmetric ones with increasing Bi concentration, which means that Bi valence band (Bi 6$sp$ band) is located in the vicinity of the Fermi level for high Bi concentrations. XPS and XAS studies of Y$_{2-x}$Bi$_x$Ru$_2$O$_7$, on the other hand, have show the metal-insulator transition is driven by varying electron correlation strength.\cite{Oh} Indeed, the peak shift of the O 1s XAS spectra in Fig. \ref{EshiftBi} (g) can be accounted for by the nonmonotonous energy shift of the O 1$s$ core level and does not imply a chemical potential shift. Also, the Ru 4$d$ $t_{2g}$ part of O 1$s$ XAS spectra shows no obvious intensity change with Bi substitution as shown in Fig. \ref{EshiftBi} (h), consistent with a fixed band filling. The line shape of the Ru 3$d$ core level changes from symmetric to asymmetric ones with increasing Bi concentration, suggesting that the metallic screening for the Ru 3$d$ core hole increases with $x$.

Figure \ref{UPSBi}(a) shows the valence-band UPS spectra of Sm$_{2-x}$Bi$_x$Ru$_2$O$_7$ near $E_F$ across the metal-insulator transition.\cite{okamoto1} The spectra have been normalized to the integrated intensity from $E_F$ to 2 eV below it, that is, to the integrated intensity of the Ru $4d$ $t_{2g}$ band. With increasing Bi concentration, the intensity around $\sim$ 1.5 eV decreases and the intensity between $E_F$ and $\sim$ 0.8 eV increases. The existence of a crossing point at $E_B$ $\sim$ 0.8 eV indicates that the spectra consist of the high binding energy component (incoherent part) and the low binding energy component (coherent part) and that there is  spectral weight transfer from the former to the latter with Bi concentration. The spectral intensity at $E_F$ increases with Bi concentration and the Fermi edge appears at $x$ $\sim$ 0.6, which resembles the spectra of Sm$_{2-x}$Ca$_x$Ru$_2$O$_7$. As shown in Fig. \ref{UPSBi} (b), spectral intensity at $E_F$ and around $E_F$ significantly change at 0.4 $\leq x \leq$ 0.6. This corresponds well to the electrical resistivity of Sm$_{2-x}$Bi$_x$Ru$_2$O$_7$ that the most dramatic change occurs around $x_t$ $\sim$ 0.6.\cite{cit2}

\subsection{Comparison between Sm$_{2-x}$Ca$_x$Ru$_2$O$_7$ and Sm$_{2-x}$Bi$_x$Ru$_2$O$_7$}
Here, we emphasize that the spectral changes of the entire Ru $4d$ $t_{2g}$ band are remarkably different on the energy scale of eV between Sm$_{2-x}$Bi$_x$Ru$_2$O$_7$ and Sm$_{2-x}$Ca$_x$Ru$_2$O$_7$. In Sm$_{2-x}$Ca$_x$Ru$_2$O$_7$, the spectra are shifted upward and some spectral weight is transferred from below $E_F$ to above it and simultaneouly the density of states (DOS) at $E_F$ increases, as expected for a filling-control Mott-Hubbard system as shown in Fig. \ref{Mott} (a). In Sm$_{2-x}$Bi$_x$Ru$_2$O$_7$, spectral weight around $E_F$ is transferred from the higher binding energy region (1$-$2 eV) to the lower binding energy region (0$-$1 eV) across the insulator-to-metal transition, as expected for a bandwidth-control Mott-Hubbard system [Fig. \ref{Mott} (b)\cite{cit4}].

As far as the spectral change in the vicinity of $E_F$ is concerned, however, both systems show similar tendency across the insulator-to-metal transitions where the spectral intensity at $E_F$ suddenly increases around $x_t$.\cite{okamoto1} The intensity at $E_F$ changes at $\sim$ $x_t$ rather suddenly in Sm$_{2-x}$Bi$_x$Ru$_2$O$_7$ whereas it changes rather gradually in Sm$_{2-x}$Ca$_x$Ru$_2$O$_7$ as can be seen in Fig. \ref{CaUPS} (b) and Fig. \ref{UPSBi} (b). That is, the intensity at $E_F$ increases by a factor of $\sim$ 10 in Sm$_{2-x}$Bi$_x$Ru$_2$O$_7$ (between $x$ = 0.4 and 0.6) while it increases by a factor of $\sim$ 5 in Sm$_{2-x}$Ca$_x$Ru$_2$O$_7$ (between $x$ = 0.2 and 0.4). This may correspond to the theoretical prediction by Watanabe and Imada\cite{Imada1} that the bandwidth-control metal-insulator transition and the filling-control metal-insulator transition have tendencies to become first-order and second-order, respectively.

According to the DMFT calculations, the intensity at $E_F$ does not change within the metallic phase for both filling-control and bandwidth-control systems as shown in Fig. \ref{Mott}. The experimetal results, however, show that the intensity at $E_F$ changes within the metallic phase for both systems [Figs. \ref{CaUPS} (b) and \ref{UPSBi} (b)]. This results in the pseudogap-like DOS around $E_F$ in the metallic phase near $x_t$. This observation may be explained by the momentum dependence of the self-energy, which is neglected in DMFT, but may be significant in real systems. On the other hand, because the increase of the incoherent part and hence the decrease of the coherent part due to surface effect have been pointed out in the studies of Ca$_x$Sr$_{1-x}$VO$_3$\cite{SCVO3S,SCVO3BS} and La$_{1-x}$Ca$_x$VO$_3$,\cite{LCVO3} the evaluation of the intensity at $E_F$ is not straightforward. Systematic bulk sensitive PES experiments using higher photon energies are necessary to clarify this point.

\section{Conclusion} 
The electronic structures of the filling-control system Sm$_{2-x}$Ca$_x$Ru$_2$O$_7$ ($0 \leq x \leq 0.6$) and the bandwidth-control system Sm$_{2-x}$Bi$_x$Ru$_2$O$_7$ ($0 \leq x \leq 2$) across the metal-insulator transition have been systematically studied using photoemission and x-ray absorption spectroscopy. In Sm$_{2-x}$Ca$_x$Ru$_2$O$_7$, we have observed monotonous energy shifts of the valence band and core levels and only small changes in the spectral shape of the Ru 4$d$ $t_{2g}$ band, indicating a nearly rigid band-like behavior of the hole doping. In Sm$_{2-x}$Bi$_x$Ru$_2$O$_7$, on the other hand, no clear energy shift was observed in most of the core levels (except for small $x$ region) but the line shape of the Ru 3$d$ core level shows increasingly metallic screening with Bi concentration. Large spectral weight transfer within the Ru $4d$ $t_{2g}$ band was observed, as expected for a bandwidth-control Mott-Hubbard system, but the partial hole transfer from Bi valence band to Ru 4$d$ $t_{2g}$ band may not be negligible. The intensity at $E_F$ shows more abrupt increase in the bandwidth-control Sm$_{2-x}$Bi$_x$Ru$_2$O$_7$ than in the filling-control Sm$_{2-x}$Ca$_x$Ru$_2$O$_7$, which may correspond to the recent theoretical prediction by Watanabe and Imada.\cite{Imada1}

\section*{Acknowledgement}
The authors would like to thank T. W. Noh for fruitful discussion. The authors would also like to thank K. Okazaki and S. Nawai for their valuable technical support in the measurements.  This work was supported by a Grant-in-Aid for Scientific Research in Priority Area "Invention of Anomalous Quantum Materials" from the Ministry of Education, Culture, Sports, Science and Technology, Japan.

\clearpage
\begin{figure}
\caption{Density of states (DOS) of the Mott-Hubbard system predicted by dynamical mean field theory\cite{Kajueter1,Zhang1}. $U$ is the on-site Coulomb energy and $D$ is the half bandwidth. Energy $\omega$ is given in units of $D$. The Fermi level is at $\omega$ = 0. (a) Filling-control system for $U$/$D$ = 4.\cite{Kajueter1} $\delta$ is the hole concentration per atom. (b) Bandwidth-control system. The critical value for MIT $U_C$/$D$ is $\simeq$ 3. The bands centered at $\omega$ = 2 and -2 are the upper and lower Hubbard bands, respectively.}
\label{Mott}
\end{figure}
\begin{figure*}
\caption{(Color) Spectra of Sm$_{2-x}$Ca$_x$Ru$_2$O$_7$ ($0 \leq x \leq 0.6$). (a) Valence band UPS spectra, (b) Ru 3$d$ core-level XPS spectra, (c) O 1$s$ core-level XPS spectra, (d) Sm 3$d_{5/2}$ core-level XPS spectra, (e) Energy shifts of the spectra plotted against Ca concentration $x$, (f) O 1$s$ XAS spectra, and (g) O 1$s$ XAS spectra normalized to the intensity in the Ru 4$d$ $e_g$ region.}
\label{CaXPS}
\end{figure*}
\begin{figure}
\caption{(Color) UPS spectra of Sm$_{2-x}$Ca$_x$Ru$_2$O$_7$ ($0 \leq x \leq 0.6$) in the Ru 4$d$ $t_{2g}$ band region. (a) Spectra across the metal-insulator transition.\cite{okamoto1} Inset shows a blow-up near $E_F$. (b) Spectral intensity at $E_F$ and that integrated within 40 meV of $E_F$ plotted against Ca concentration $x$.}
\label{CaUPS}
\end{figure}
\begin{figure}
\caption{(Color) UPS spectra of Sm$_{2-x}$Ca$_x$Ru$_2$O$_7$ ($0 \leq x \leq 0.6$) arbitrarily shifted so that the intensity at energy zero coincides.}
\label{RefShift}
\end{figure}
\begin{figure*}
\caption{(Color) Spectra of Sm$_{2-x}$Bi$_x$Ru$_2$O$_7$ ($0 \leq x \leq 2$). (a) Valence-band UPS spectra,  (b) Ru 3$d$ core-level XPS spectra, (c) O 1$s$ core-level XPS spectra, (d) Sm 3$d_{5/2}$ core-level XPS spectra, (e) Bi 4$f$ core-level XPS spectra, (f) Energy shifts of these spectra plotted against Bi concentration $x$, (g) O 1$s$ XAS spectra, and (h) O 1$s$ XAS spectra normalized to the intensity in the Ru 4$d$ $e_g$ region.}
\label{EshiftBi}
\end{figure*}
\begin{figure}
\caption{(Color) UPS spectra of Sm$_{2-x}$Bi$_x$Ru$_2$O$_7$ ($0 \leq x \leq 2$) in the Ru 4$d$ $t_{2g}$ band region. (a) Spectra across the metal-insulator transition.\cite{okamoto1} Inset shows a blow-up near $E_F$. (b) Spectral intensity at $E_F$ and that integrated within $\pm$ 40 meV of $E_F$ plotted against Bi concentration $x$.}
\label{UPSBi}
\end{figure}



\begin{thebibliography}{99}
\bibitem[*]{address1} Present Address: National Synchrotron Radiation Research Center, Hsinchu 30077, Taiwan R. O. C.

\bibitem[**]{Yaddress1} Present Address: Research Center for Materials Science at Extreme Condition, Osaka University, Toyonaka, Osaka 560-8531, Japan

\bibitem{cit4} M. Imada, A. Fujimori, and Y. Tokura, Rev. Mod. Phys. \textbf{70}, 1039 (1998).

\bibitem{Zhang1} X. Y. Zhang, M. J. Rozenberg and G. Kotliar, Phys. Rev. Lett. {\bf 70}, 1666 (1993).

\bibitem{Kajueter1} H. Kajueter, G. Kotliar and G. Moeller, Phys. Rev. Lett. {\bf 53}, 16214 (1996).

\bibitem{SCVO3S} I. H. Inoue, I. Hase, Y. Aiura, A. Fujimori, Y. Haruyama, T. Maruyama and Y. Nishihara, Phys. Rev. Lett. {\bf 74}, 2539 (1995).

\bibitem{SCVO3B} R. Eguchi, T. Kiss, S. Tsuda, T. Shimojima, T. Yokoya, A. Chainani, S. Shin, I. H. Inoue, T. Togashi, S. Watanabe, C. Q. Zhang, C. T. Chen, M. Arita, K. Shimada, H. Namatame, and M. Taniguchi, cond-mat/0504576.

\bibitem{CSRO3T} M. Takizawa, D. Toyota, H. Wadati, A. Chikamatsu, H. Kumigashira, A. Fujimori, M. Oshima, Z. Fang, M. Lippmaa, M. Kawasaki, and H. Koinuma, cond-mat/0417218.

\bibitem{LSTO3} A. Fujimori, I. Hase, M Nakamura, H. Namatame, Y. Fujishima, Y. Tokura, M. Abbate, F. M. F. de Groot, M. T. Czyzyk, J. C. Fuggle, O. Strebel, F. Lopez, M. Domke and G. Kaindl, Phys. Rev. B {\bf 46}, 9841 (1992).

\bibitem{Fujimori1} A. Fujimori, I. Hase, H. Namatame, Y. Fujishima, Y. Tokura, H. Eisaki, S. Uchida, K. Takegahara and F. M. F. de Groot, Phys. Rev. Lett. {\bf 69}, 1796 (1992).

\bibitem{LSCO4} A. Ino, T. Mizokawa, K. Kobayashi, A. Fujimori, T. Sasagawa, T. Kimura, K. Kishio, K. Tamasaku, H. Eisaki and S. Uchida, Phys. Rev. Lett. {\bf 81}, 2124 (1998).

\bibitem{cit3} S. Yoshii, K. Murata and M. Sato, J. of Phys. Chem. Solids {\bf 62}, 129 (2001).

\bibitem{cit1} T. Yamamoto, R. Kanno, Y. Takeda, O. Yamamoto, Y. Kawamoto, 
and M. Takano, J. of Solid State Chem. {\bf 109}, 372 (1994).

\bibitem{cit2} S. Yoshii, K. Murata, and M. Sato, J. of Phys. Soc. Jpn. {\bf 69}, 17 (2000).

%
\bibitem{cit0} J. B. Goodenough, A. Hamnett, and D. Tells, 
{\it Localization and Metal-Insulator Transition}, Plenum, New York 1985, p.161. 

\bibitem{okamoto1} J. Okamoto, S.-I. Fujimori, T. Okane, A. Fujimori, M. Abbate, S. Yoshii, and M. Sato, Acta Physica Polonica B {\bf 34}, 783 (2003).

\bibitem{Tl2Ru2O7} J. Okamoto, T. Mizokawa, A. Fujimori, T. Takeda, R. Kanno, F. Ishii, and T. Oguchi, Phys. Rev. B {\bf 69}, 35115 (2004).

\bibitem{Noh} J. S. Lee, S. J. Moon, T. W. Noh, T. Takeda, R. Kanno, S. Yoshii, and M. Sato, Phys. Rev. B, in press.

\bibitem{Oh} J. Park, K. H. Kim, H.-J. Noh, S.-J. Oh, J.-H. Park, H.-J. Lin, and C.-T. Chen, Phys. Rev. B {\bf 69}, 165120 (2004).

\bibitem{Imada1} S. Watanabe and M. Imada, J. Phys. Soc. Jpn. {\bf 73}, 1251 (2004).


\bibitem{SCVO3BS} A. Sekiyama, H. Fujiwara, S. Imada, S. Suga, H. Eisaki, S. I. Uchida, K. Takegahara, H. Harima, Y. Saitoh, I. A. Nekrasov, G. Keller, D. E. Kondakov, A. V. Kozhenvnikov, TH. Pruschke, K. Held, D. Vollhardt and V. I. Anisimov, Phys. Rev. Lett. {\bf 93}, 156402 (2004).


\bibitem{LCVO3} K. Maiti, P. Mahadevan and D. D. Sarma, Phys. Rev. Lett. {\bf 80}, 2885 (1998): K. Maiti and D. D. Sarma, Phys. Rev. B {\bf 61}, 2525 (2000). 

\end{thebibliography}
\end{document}